\documentclass[12pt,preprint]{aastex}
\newcommand{\pref}{\protect\ref}
\newcommand{\solrad}{\ifmmode{R}_{\rm S}\else${R}_{\rm S}$\fi}
\newcommand{\solmas}{\ifmmode{M}_{\rm S}\else${M}_{\rm S}$\fi}

\newcommand{\ctn}{\ifmmode\kappa\else$\kappa$\fi}

\newcommand{\velu}{$\,$km$\,$s$^{-1}$}
\newcommand{\dynu}{$\,$dyn$\,$cm$^{-2}$}
\newcommand{\term}[2]{\mbox{$\,^{#1}{\rm #2}$}}
\def\term#1 #2/{\mbox{$\,^{#1}{\rm #2}$}}

\def\mathstacksym#1#2#3#4#5{\def#1{\mathrel{\hbox to 0pt{\lower 
    #5\hbox{#3}\hss} \raise #4\hbox{#2}}}}

\mathstacksym\lta{$<$}{$\sim$}{1.5pt}{3.5pt} 
\mathstacksym\gta{$>$}{$\sim$}{1.5pt}{3.5pt} 
\mathstacksym\lrarrow{$\leftarrow$}{$\rightarrow$}{2pt}{1pt} 
\mathstacksym\lessgreat{$>$}{$<$}{3pt}{3pt} 

\newcommand\figone{
\begin{figure}[!ht] 
\epsscale{0.9}
\plotone{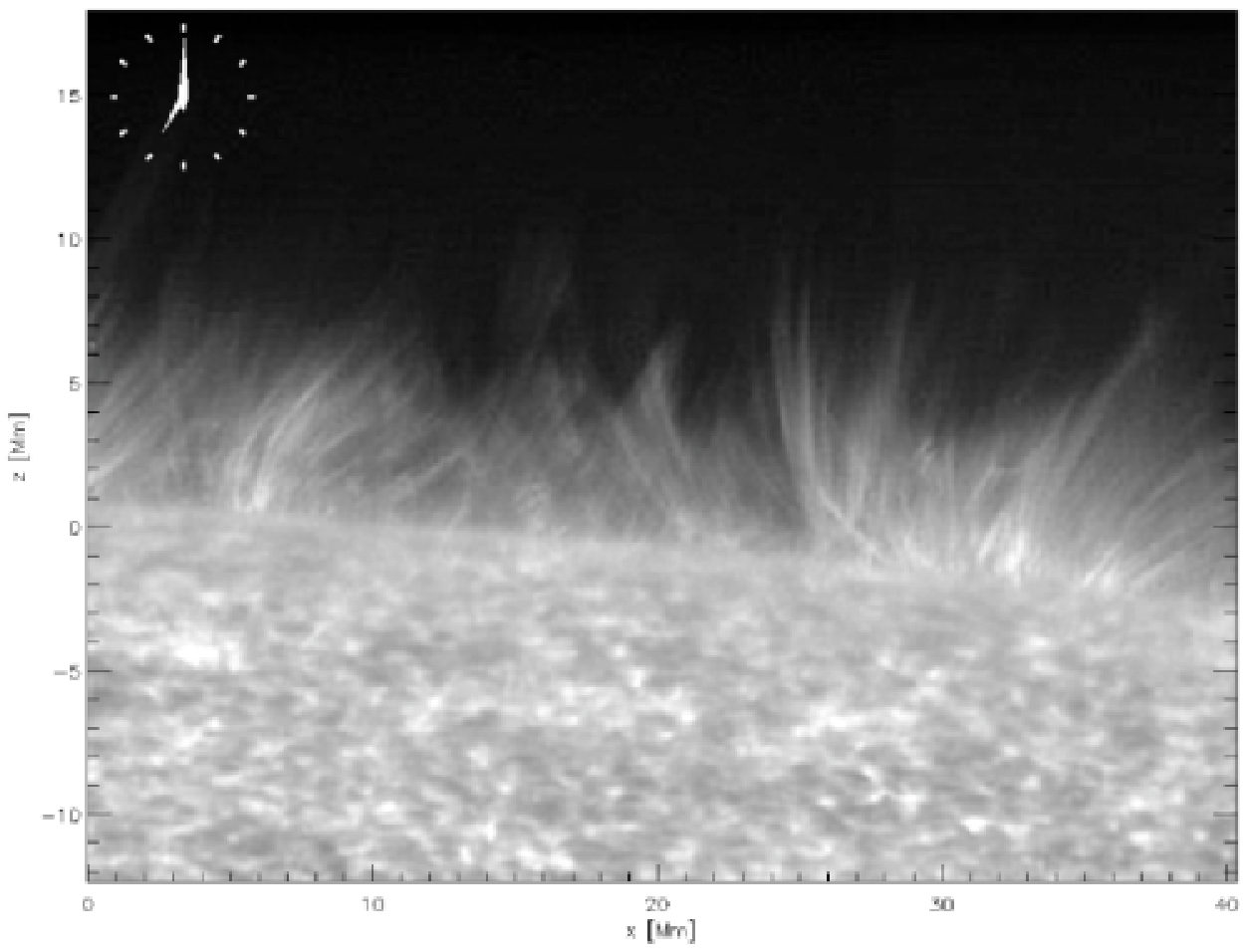}  
\caption{\label{fig:snapshot} 
A snapshot from a time series of Ca~II images obtained with the 3 \AA{}
wide filter of the BFI instrument using the SOT on the Hinode
spacecraft.  The height scale has been carefully determined, relative
to
the vertical continuum (5000 \AA) optical depth unity, by Bj\o{}lseth. 
}
\end{figure}
}

\newcommand\figtwo{
\begin{figure}[!ht] 
\epsscale{0.9}
\plotone{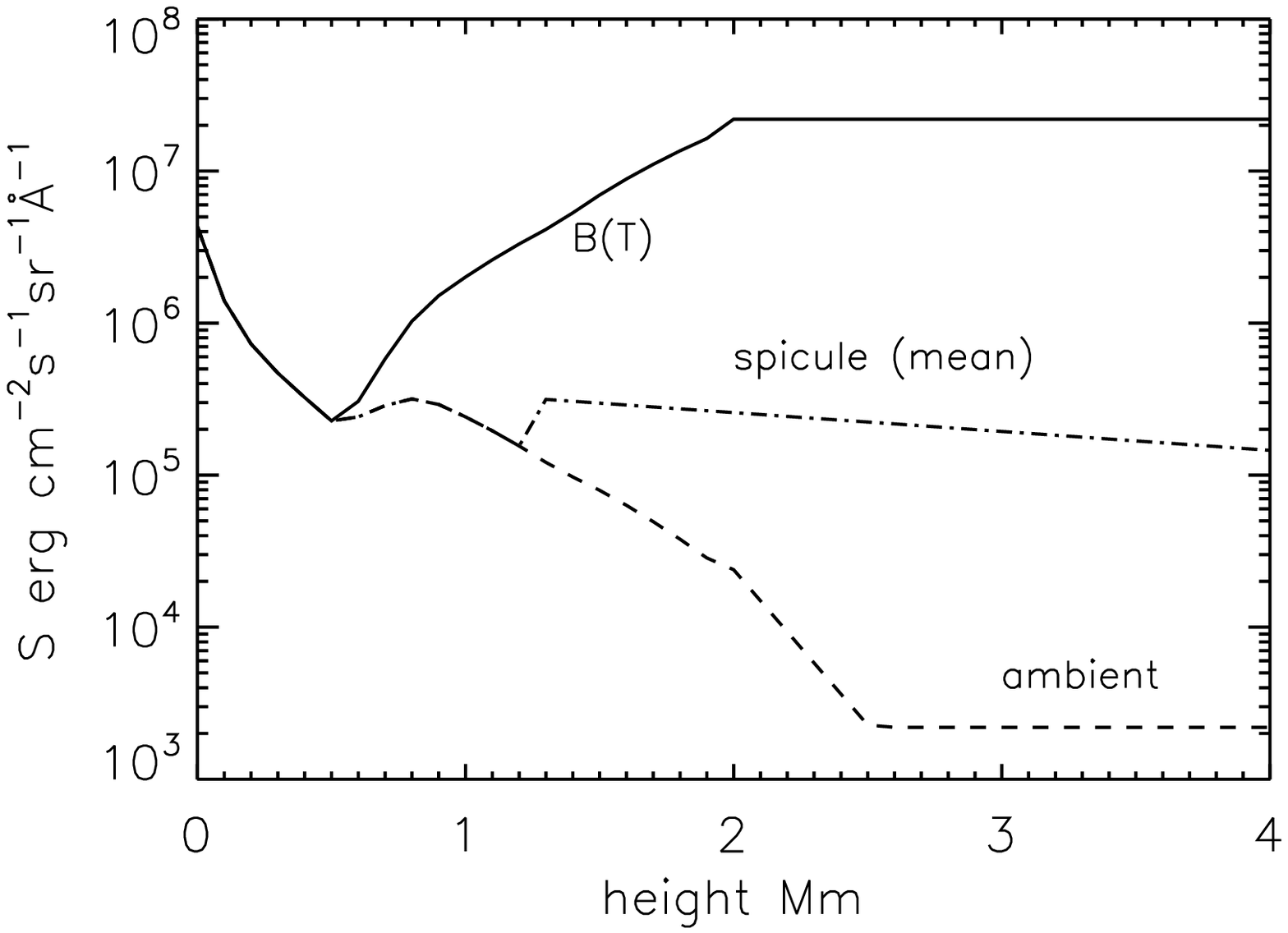}  
\caption{\label{fig:s} 
Prescribed source functions used in the calculations.  The spicules
were treated statistically. 
}
\end{figure}
}

\newcommand\figthree{
\begin{figure}[!ht] 
\epsscale{0.9}
\plotone{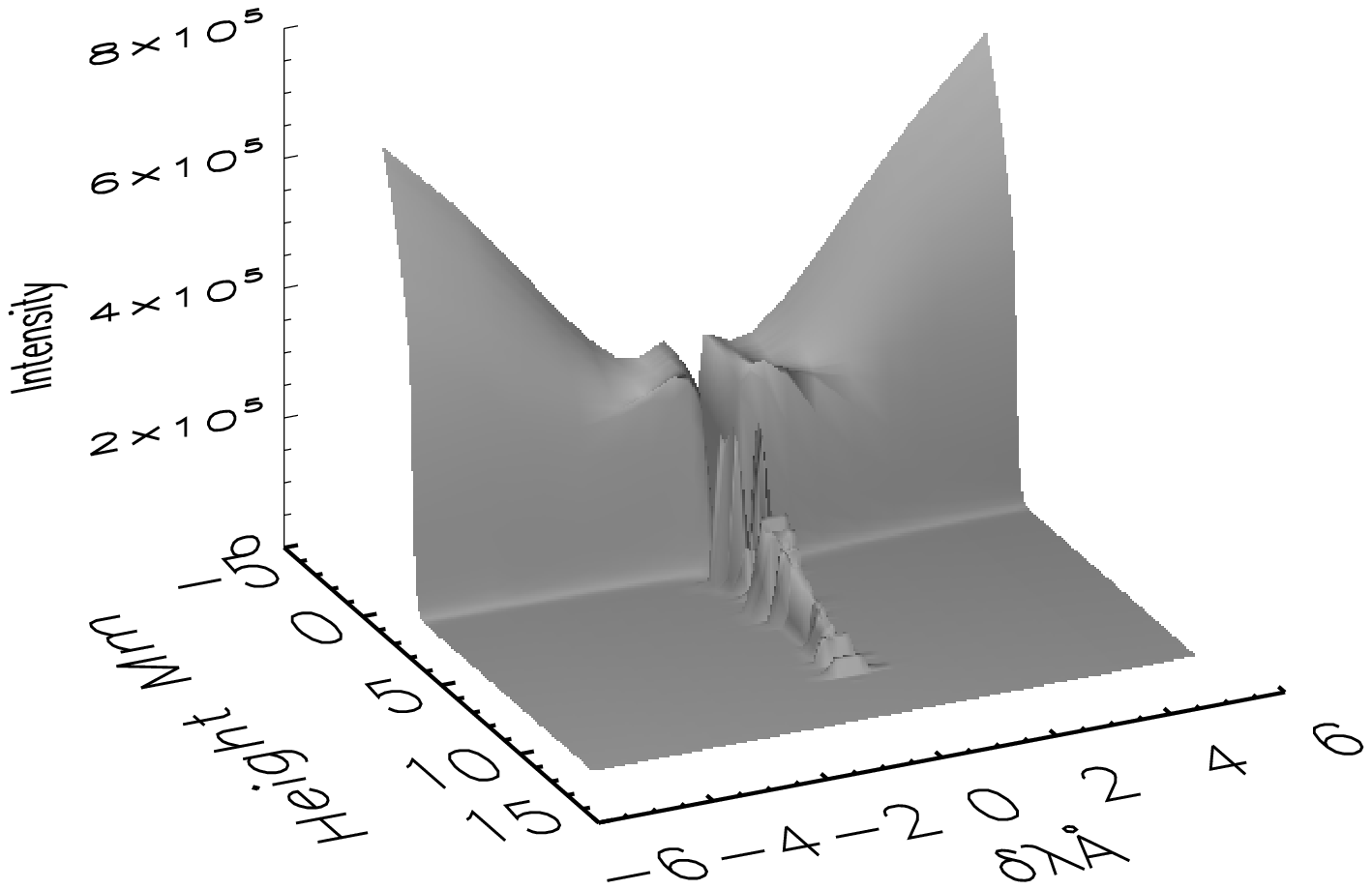}  
\caption{\label{fig:profiles} 
Line profiles computed from a radial slice of the 
``standard'' transfer calculations,
simply to demonstrate that the computed spectra are not dissimilar to
those observed. The self-reversed core in the photosphere (heights
near zero) change to emission in the spicular material at greater heights.
}
\end{figure}
}

\newcommand\figfour{
\begin{figure}[!ht] 
\epsscale{1.0}
\plottwo{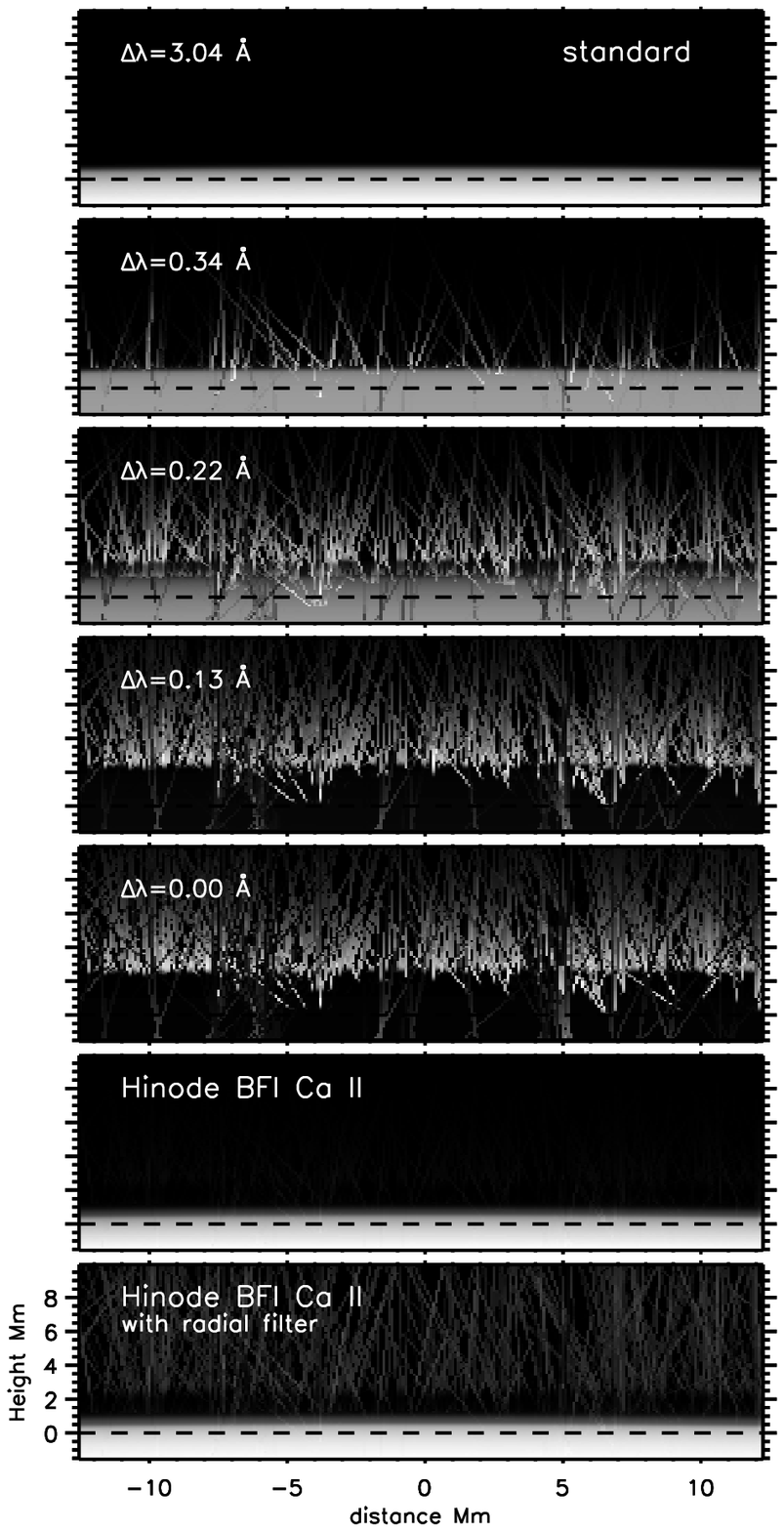}{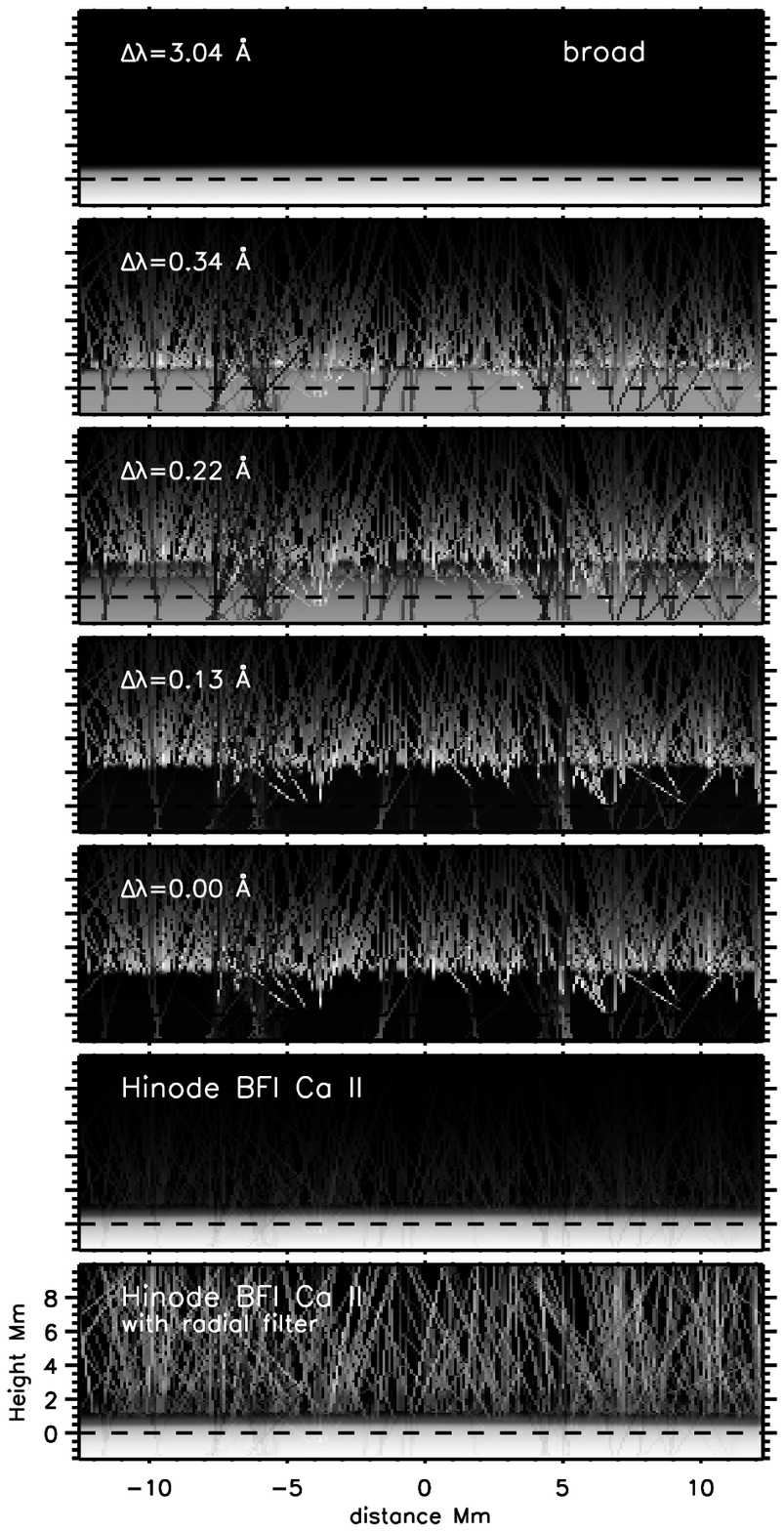}  
\caption{\label{fig:standard} 
Intensities computed at several monochromatic wavelengths and in the 
 Hinode BFI passband are shown 
as a function of position along the limb tangential diretion and radial height.
The ``standard'' spicule conditions were applied (left panel), and
broad spicular emission lines were computed (right panel). 
}
\end{figure}
}

\newcommand\figfive{
\begin{figure}[!ht] 
\epsscale{0.8}
\plotone{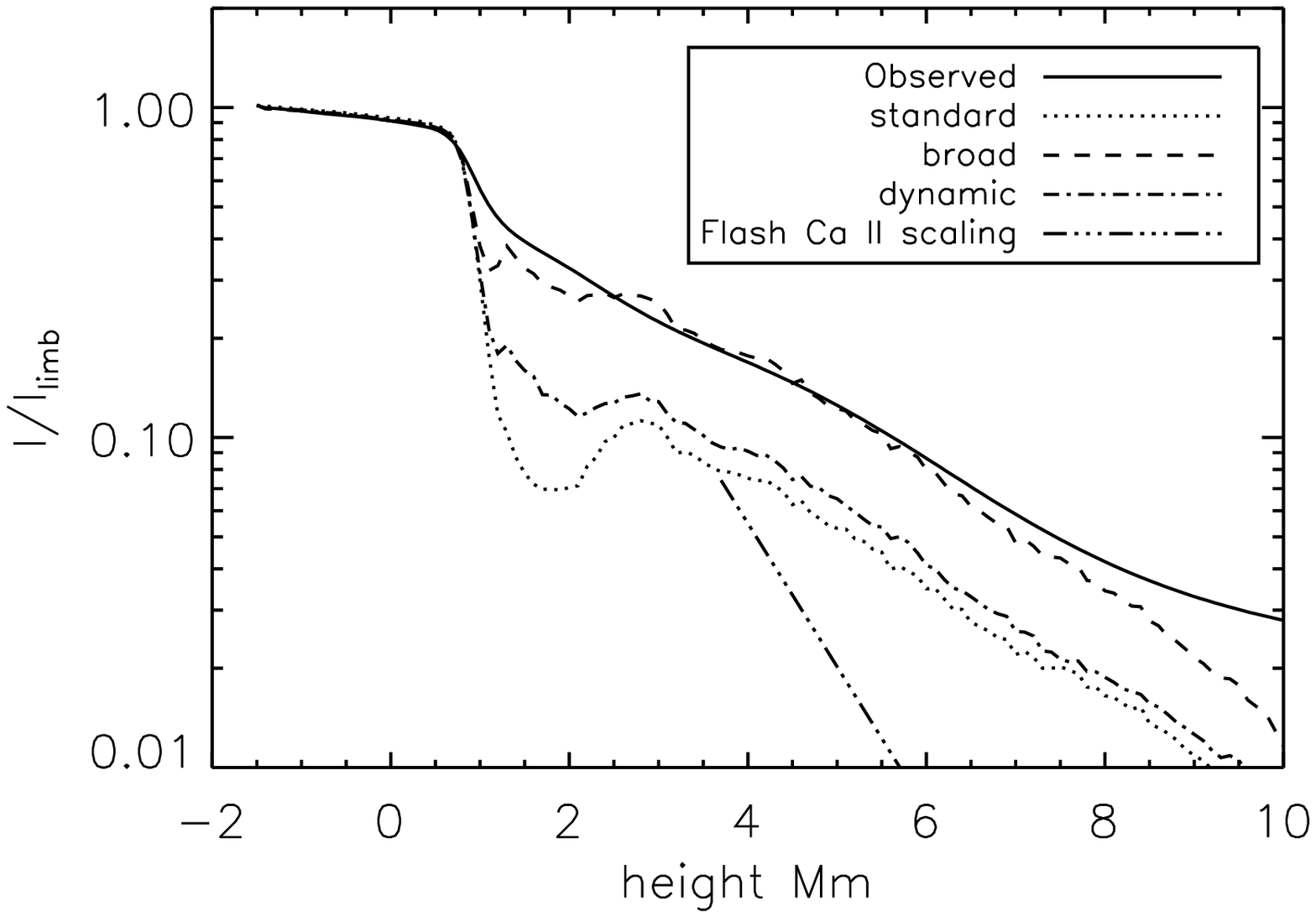}
\caption{\label{fig:compare} 
Average intensities, normalized to the same point within the solar
limb, are plotted for observations and for three model calculations.  
Also shown is the height dependence of the Ca~II emission found by
\protect\citet{Makita2003} from eclipse flash spectra, the absolute
value of which is arbitrary.
}
\end{figure}
}

\slugcomment{}

\begin{document}
\title{On the solar chromosphere observed at the limb with Hinode}

\author{Philip G. Judge}
\affil{High Altitude Observatory,
National Center for Atmospheric Research\altaffilmark{1},
P.O. Box 3000, Boulder CO~80307-3000, USA\\ \vbox{}}

\and

\author{Mats Carlsson}
\affil{ Institute of Theoretical Astrophysics,
           P.O.~Box~1029, Blindern, N--0315 Oslo, Norway}

\altaffiltext{1}{The National Center for Atmospheric Research is
sponsored by the National Science Foundation}

\begin{abstract}
  Broad-band images in the Ca~II H line, from the BFI instrument 
  on the Hinode spacecraft,
  show emission from spicules emerging from and visible right down to
  the observed limb.  Surprisingly, little absorption of 
  spicule light is seen
  along their lengths.  We present formal solutions to the transfer
  equation for given (ad-hoc) source functions, including a
  stratified chromosphere from which spicules emanate.  The model
  parameters are broadly compatible with earlier studies of spicules.  
  The visibility of Ca~II 
  spicules down to the limb in Hinode data 
  seems to require that spicule emission  be 
  Doppler shifted relative to  the stratified atmosphere, either by
  supersonic turbulent or organized spicular motion.  The  non-spicule
  component of the chromosphere is almost invisible in the
  broad band BFI data, but we predict that it will be clearly visible in high
  spectral resolution data.  Broad band Ca II H limb images give the false
  impression that the chromosphere is dominated by spicules.  
  Our analysis serves as a reminder that
  the absence of a signature can be as significant as its presence.
\end{abstract}

,

\section{Introduction}

The Hinode spacecraft is a stable platform from which unique high
resolution, seeing-free images of the Sun can be acquired
 \citep{Kosugi+others2007}. The BFI instrument, fed by the
Solar Optical Telescope (SOT) on Hinode \citep{Tsuneta+others2008},
can observe a 3 \AA{} wide spectral bandpass centered at
the H line of Ca~II. Over this bandpass, the line 
forms in both the photosphere in the wings, and chromosphere 
in the core. 
Movies of such Ca II images 
have revealed a remarkably dynamic,
spicule-dominated limb. 
The observed spicules have smaller diameters, 
higher apparent velocities and smaller
lifetimes \citep{DePontieu+others2007} than was previously thought 
\citep{Beckers1968,Beckers1972}.   

Figure~\pref{fig:snapshot} shows a typical snapshot from a series of
Ca~II BFI images acquired on 7 November 2007, in the northern polar
coronal hole.    We selected a
coronal hole because spicules there are longer than elsewhere, thereby
providing a broad background of spicule emission against which a
stratified atmosphere might easily be identified. 
A smooth radial gradient has been divided out of the
data to enhance the emission high above the limb.  
The zero point of the 
height scale $z=0$ (along $x=0$) corresponds to 
the standard formation
height of the 5000 \AA{} continuum, when observed vertically, as 
derived by \citet{Bjolseth2008}.  
Henceforth we will refer to heights on this standard scale.
Bj\o{}lseth found that 
the Ca~II limb lies $0.45\pm0.034$  Mm above the 
blue limb.    Since the 
continuum at the limb forms about 0.375 Mm higher than at disk center, 
(e.g., \citealp{Athay1976}), 
the Ca II limb forms near heights of 0.825 Mm. 

Curiously, such limb images show little or no signature
of a bulk, stratified chromosphere\footnote{By ``chromosphere'' we
  refer not only to the traditional definition of H$\alpha$ emitting
  plasma seen during eclipse flashes, but all the material lying
  between the quiet photosphere, with densities above $\sim
  5\times10^{-9}$ g~cm$^{-3}$, and the corona with densities below
  $\sim 10^{-13}$ g~cm$^{-3}$.  } which, as we discuss below, should
have a thickness between 1 and 2 Mm.  The BFI instrument's resolution
($\sim 0\farcs1$) is ample to resolve structure on scales of 1-2 Mm
($0\farcs1 \equiv 0.0725$ Mm).  Yet a striking feature of these images
is the continuous emission seen along each spicule all the way down
to, and sometimes across, the limb.  Where then is the stratified
chromosphere, and why are spicules so obviously dominant that one
might conclude that the chromosphere itself consists of little more
than a collection of spicules?
In this paper we explore what these observations imply
in terms of the structure of the chromosphere.

\section{Simple calculations}
\label{sec:calculations}

The complex dynamic behavior of the Ca~II spicules, their unknown
origin and other difficulties preclude the possibility of meaningful
ab-initio or other detailed modeling efforts.  To address the above
questions, 
a far simpler calculation is appropriate.  
We model the chromosphere 
as a stratified atmosphere from which spicules  
emanate.  Formal solutions to the transfer
equation along rays tangential to the solar limb are performed for
prescribed source functions, densities and atomic parameters
applicable to the Ca~II H line.  

Our adopted stratified medium is simply the 1D atmosphere of
\citet{Gingerich+others1971}. As suggested by the Hinode BFI
observations, this medium does not emit as much as the embedded
spicules, so it serves primarily to scatter photons.  
The exact stratification of the bulk
chromosphere
is not critical, all
that is required is that it span the range of densities from
photosphere to corona in $\lta 2$ Mm.  The mean stratifications of hydrodynamic or
MHD models, such as by \citet{Carlsson+Stein1995,Wedemeyer+Lagg+Nordlund2009}, are
similar to the stratification used here.

The stratified atmosphere was assigned source functions of $B_\nu(T(r))f(r)$
where $T(r)$ is taken from the atmospheric model, as a function of
radial distance from Sun center $r$.  $B_\nu(T)$ is the Planck
function at frequency $\nu$ and temperature $T$.  The function
\footnote{Here $f(r)=1$ ($r<r_1$) and $e^{-(r-r_1)/1.7\ell}$ where
  $r_1=r_0+4.4\ell$, $\ell=0.125$ Mm being a typical pressure scale
  height in the photosphere.}  $f(r)$ mimics well-known
non-LTE effects in which source
functions fall below their LTE values.  Figure~\pref{fig:s} shows the
source functions used in our calculations.

The spicules were treated statistically, both in their spatial
distributions and thermodynamic properties.  They were randomly
distributed along the boundaries of circular supergranules into 200
small bushes with a common ``root'', 8 spicules in each bush.  1600
spicules per supergranule, each with a diameter of 0.1 Mm, leads to a
filling factor by area of 0.015, and a total of $2\times10^7$ such
spicules on the Sun at any time. (The latter is some 20 times larger
than the value derived by Beckers 1972,  based upon 
data of far lower angular resolution than those from the Hinode BFI instrument). 
These numbers produce synthetic spicule images
similar to those from Hinode.

The roots of individual spicules are set at $r_0=1.25$ Mm above the
continuum photosphere, with a base density $\rho=\rho_0=1.4\times 10^{-11}$
g~cm$^{-3}$. 
Below the $r_0$ there are no spicules in
our calculations- the atmosphere is 100\% ``stratified''.  The
spicules are modeled simply as longer extensions of the atmosphere
from this base.  
The spicule densities are
$\rho(r)=\rho_0\,e^{-(r-r_0)/\ell_s}$ with $\ell_s=3.5$ Mm for
$r>r_0$.  The scale height $\ell$ was chosen to match intensity scale
heights of $\sim 3.5$ Mm found for polar coronal holes by
\citep{Bjolseth2008}, in order to compare calculations with
Figure~\pref{fig:snapshot}.  The calculated intensities depend only
weakly on the gas densities, because our source functions are fixed
and the spicules have optical depths of order $10^1$ in the line
cores.  The assumed spicule properties will be revisited in the 
Section \pref{sec:discussion}.

Spicule orientations were randomized relative to the local
vertical in azimuth, and their source functions specified along each one's length
but randomly varied between spicules.  The source functions are not
individually known, being determined locally by collisional excitation
and scattering of radiation from the bright underlying photosphere.
\citet[][fig. 12]{Makita2003} shows source functions below 4 Mm with
black body temperatures near 4300K.  Here, each spicule's base source
function was chosen from a randomly distributed sample about a mean of
$B_\nu(T=4300 {\rm K})$ with an arbitrary  distribution width one tenth of this.
Along each spicule the source function drops with height along
with the density.  Fig.~\pref{fig:s} shows the mean value as a
function of height.  For the line opacity, all calculations use a
calcium logarithmic abundance of 6.3 (where H=12), all calcium is
assumed to be in Ca~II (a good approximation below the transition
region) an absorption oscillator strength of 0.33, a
microturbulence of 10 km~s$^{-1}$ (except where specified below) and a
radiatively damped Voigt profile.  Standard continuum opacity was
added to the line opacity from \citet{Allen1973}.

\section{Results}

Figure~\pref{fig:profiles} shows intensity profiles of the H line as a
function of wavelength and height in a ``standard'' model.  
The computed intensities are similar to
observations both above and below the limb.  The chosen Ca II line
parameters produce results not incompatible with
typical profiles seen on the disk \citep{Linsky+Avrett1970}, and with
Ca~II observations obtained both during and outside eclipses
\citep{Beckers1968,Beckers1972,Makita2003}.

Figure~\pref{fig:standard} shows emergent intensities at various
monochromatic wavelengths and integrated over the BFI filter bandpass.
In the ``standard'' calculation (left
panels), the spicule line profiles are assumed to be the same
as in the stratified atmosphere.  The latter leads to absorption
at wavelengths within 0.1 \AA{} of line center and below heights of
$\sim 2$ Mm.  The path lengths and line opacity of the stratified  material
are sufficient to produce absorption, unless a spicule happens to lie
physically closer to the observer.  The limb in the simulated BFI data
(see the panel labeled ``Hinode BFI Ca~II'') is close to the $0.875$
 Mm value derived observationally by \citet{Bjolseth2008}.  In the
lowest panels, the same radial function was applied to the simulated
data as the observations shown in Figure~\pref{fig:snapshot}, to
enhance the visibility of spicule emission over the limb.  In the
broad BFI bandpass, a significant and observable fraction of spicule
emission is absorbed by the stratified atmosphere below 2-3 Mm.  This
behavior is inconsistent with the appearance of Hinode data.

Real spicules are dynamic, as seen both through linewidths and
physical motions
\citep{Beckers1968,Beckers1972,Makita2003,DePontieu+others2007}.
Therefore we made two further calculations: one using broad spicule
emission line profiles, and another using spicule-aligned flows.  Both
calculations  shift the spicule emission outside of 
the absorption profiles of the stratified
atmosphere  when the Doppler speeds exceed
$ \xi \sqrt{\ln\tau_0}$, where $\xi$ is the chromospheric
microturbulence ($\lta c_s$, the sound speed, $c_s \sim 7$ \velu{},
e.g., \citealp{Vernazza+Avrett+Loeser1981}),
and $\tau_0>1$ is the line center optical depth tangential to the limb.
For values of $\tau_0$ varying between 10 and $10^{10}$ 
the required shifts are a few times $c_s$.  

The right hand panels of Figure~\pref{fig:standard}
present calculations including a spicule line broadening
microturbulent parameter drawn from a distribution with a mean 
of 30 \velu{} and a standard deviation of 10 \velu{}.  
This supersonic 
microturbulence
is compatible with
linewidths measured from eclipse data below about 4 Mm
\citep[e.g.][]{Makita2003}.  (In the dynamic calculation, not shown, 
spicule-aligned outflows were drawn from a
distribution with
a mean of 120
\velu{} and a standard deviation of 40 \velu.)
In this calculation, the spicules can be
seen down to and crossing the limb, as observed, and the dark
absorption band resulting from the stratified atmospheric absorption is
less pronounced.  Figure~\pref{fig:compare} shows  intensities
in the BFI bandpass
averaged along the direction tangential to the limb, 
normalized to limb values, as a function of
height, from the three calculations and from 
observations. The
filter-integrated emission from broad spicule line profiles is larger
than from the standard calculation, because the computed individual
spicules are optically thick across their axes, at least for heights 
below 4 Mm.

The differences in the broad dips in intensity between heights of 1
and 2.5 Mm shows our main result- dynamical calculations are needed
to avoid such a large dip in BFI Ca~II intensities across the limb.

\section{Discussion}
\label{sec:discussion}

It seems that the absence of the stratified chromosphere in the images
obtained with the Hinode BFI Ca~II filter may be explained, at least
in part, simply by large Doppler shifts resulting from spicule
dynamics.  The computations (Figure~\pref{fig:standard}) 
resemble observations (Figure~\pref{fig:snapshot}) when spicule
emission is Doppler shifted out of the dark core of the H line in the
stratified chromosphere.  

\subsection{From photosphere to corona}

The solar atmosphere does not end at the visible photosphere- there
must exist material as the upwardly stratified extension of the
photosphere.  This material must, on average, be highly stratified
because quiet Sun coronal pressures are close to 0.1 \dynu{}
\citep[e.g.][]{Mariska1992}, yet photospheric pressures are orders of
magnitude higher.  The only question of interest here is if
this material is expected to be able to scatter the Ca II resonance
lines.  Even if the chromosphere were in hydrostatic and 
radiative equilibrium, and
hence maximally stratified, almost all of the calcium would be in the
Ca II ground state, and the stratified layer would span $\sim1$
Mm before coronal conditions were reached.
In semi-empirical 1D models the transition from photosphere to corona
spans 1.5 Mm (measured from temperature minimum to corona,
\citealp{Gingerich+others1971,Athay1976,Vernazza+Avrett+Loeser1981}).

This transition must, on average, be stratified close to hydrostatic equilibrium,
because motions observed in spectral lines formed in the photosphere
and chromosphere are, statistically speaking, sub-sonic.  Indeed 
one has to look hard to find the on-disk counterparts of the
highly supersonic type II spicules 
\citep{McIntosh+dePontieu2010},
for example.
More directly observable signatures of the stratified
chromospheric medium are found, for example, in the ``flash spectrum''
seen during eclipses \citep[][and references therein]{Makita2003}, 
or in the upward extension of photospheric wave
motions seen on the solar disk.  While the observationally-defined 
``chromospheric extent'' inferred by flash spectra exceeds 
hydrostatic values, it is also compatible with a hydrostatic
stratification in the first 1-1.5 Mm.  The large extent arises
primarily from the data seen 
high above the limb which are dominated by 
spicules.  
These and other issues are reviewed
by, e.g., \citet{Gibson1973,Athay1976,Judge2006}.

\subsection{Validity of our results}

The {\em ad-hoc} parameters in our calculations clearly limit their
usefulness. But our essential result- the need to Doppler shift
spicule material out of the absorbing stratified chromosphere in order
to reproduce qualitatively Hinode Ca II data- is relatively
insensitive to such details.  The result simply requires spicules to
originate close to the base of the chromosphere, and have different
source functions and/or opacities from the stratified atmosphere.
Given these conditions, and spicule lengths which exceed the thickness
of the stratified atmosphere, our result appears robust.  
Our particular choice of parameters were taken from
observed properties discussed by
\citet{Makita2003,DePontieu+others2007}.  For a given density, the opacity
in the Ca II H line follows from atomic data, ion abundance, and
thermal and non-thermal motions.  Average densities of the stratified
medium are, as we argued above, approximately in hydrostatic equilibrium.
However, our spicule densities and their height dependence were chosen
simply to produce computations qualitatively similar to the particular
coronal hole data shown in Figure~\pref{fig:snapshot}.

Physical considerations suggest that $\rho_0$ 
cannot greatly exceed $10^{-11}$ g~cm$^{-3}$.
\citet{DePontieu+others2007,McIntosh+dePontieu2010} find these
spicules to be highly supersonic, $\sim 100$ km~s$^{-1}$.  Such high
speeds require magnetic forces in plasma where the sound speed is
certainly $\lta 10$ km~s$^{-1}$.  While the mechanism driving spicules
is not known, the characteristic Alfv\'en speed $v_A$ must exceed 100
km~s$^{-1}$.  Using an upper limit of 1 kG for field strengths in the
low chromospheric network (1 kG is characteristic of network
photospheric fields), $v_A \sim 100$ km~s$^{-1}$,
we find $\rho < 10^{-9}$ g~cm$^{-3}$.  But this is an unrealistically 
large estimate, since chromospheric magnetic fields are weaker 
due to geometric expansion of network fields, and not all of the local magnetic
energy
is free to be converted to kinetic energy.  
Strong network magnetic fields tend also
to be largely unipolar, thus only tangential components associated
with magnetic shear or with weaker neighboring opposite polarity
fields
contain the free energy.  
Adopting field strengths nearer to 0.1 kG, as an order of
magnitude estimate, the observed spicule speeds require $\rho \sim
10^{-11}$ g~cm$^{-3}$, as used above.  It is difficult to see how this
estimate can be significantly larger.

Spicules in coronal holes are longer, as seen in ground based data
\citep{Beckers1972} and in Hinode data
\citep{DePontieu+others2007,Bjolseth2008}.  Our scale height of 3.5 Mm
for coronal hole densities and source functions is twice the value
derived for the numbers of spicules observed as a function of height
for the Sun in general by \citet{Beckers1972}.  \citet{Bjolseth2008}
shows in her Fig.~4.10 that the Hinode Ca II data of equatorial
regions have scale heights closer to 2 Mm.  The relationship between
the spicules observed by the Hinode BFI instrument and earlier work
has not yet been clarified.  We simply note that our calculations are
not unrealistic parameterizations of the conditions needed to describe
radiative transfer in the chromosphere of a coronal hole.

Our computations are not in qualitative disagreement with the 1.5 Mm
wide dip in H$\alpha$ line center intensities discovered by
\citet{Loughhead1969}.  The cores of H$\alpha$ and neutral helium
lines routinely show a dip in intensity surrounded by a shell of
emission \citep[e.g.][and much later
  work]{White1963,Loughhead1969,Pope+Schoolman1975}.  
However, dips seen in visible lines of hydrogen and helium may result
more from the well-known lack of opacity in the low to mid
chromosphere, and extra opacity due to fibrils which appear to
over-arch the stratified chromosphere.  Resolving the issue would
require detailed calculations of H$\alpha$ and He lines with models
taking into account the fibril structure and excitation mechanisms populating 
these excited atomic levels, which are not
currently feasible.

Lastly, there remains an interesting discrepancy between the Hinode BFI
data and flash spectra, in that the 2 Mm Ca II scale height is twice
the median value derived from flash spectra of Ca II lines
\citep[][his table 2, see also our
  Fig.~\pref{fig:compare}]{Makita2003}.  Importantly we also note that
our calculations never remove the off-limb dip entirely, even in the
dynamic and broad line calculations, yet at least some of the Hinode
BFI images appear to show no hint of a dip.

\subsection{Implications}

Both observations and simple physical arguments require
that spicules 
be a consequence of some plasma or
magnetohydrodynamic processes occurring  within the 
chromosphere.  Spicules cannot arise fully fledged from the
photosphere for several reasons, not least because photospheric and
spicular densities and gas pressures differ by many orders
of magnitude. 

Our result suggests that
the Hinode BFI Ca II images can be used as speedometers in the sense
that spicules, when visible down to the limb, must have components that are
Doppler shifted supersonically, say by more that say 20 \velu{}.  
While the limb chromosphere
appears in the Hinode BFI Ca~II data to be made entirely of spicules,
this broad bandpass appears to be almost blind to much of the stratified,
inter-spicule chromosphere.  The Hinode BFI Ca II filter probably
misses populations of the short ``type-I''  spicules also associated with
the magnetic network,  whose line widths and
Doppler shifts are insufficient to avoid the absorption by the
intervening material. 

In fact, these Hinode data miss the bulk of the
mass of the chromosphere, including the internetwork. 
Standard chromospheric models give on average 0.03 g~cm$^{-2}$ as the
total chromospheric surface mass density
\citep[e.g.][]{Vernazza+Avrett+Loeser1981}.  The mass density per unit
area of
the spicules, averaged over the surface, is $\sim \rho \ell f$ where
$\rho$, $\ell$ and $f$ are their typical mass density, length and
surface filling factor.  Using $\rho \sim 1.4\times 10^{-11}$
g~cm$^{-3}$, $\ell \sim 3.5\times10^8$ cm, $f=0.015$, we find an
average mass density of only $7\times10^{-5}$ g~cm$^{-2}$.  The
spicules observed by the Hinode Ca~II BFI instrument comprise less
than 0.3\% of the entire chromospheric mass. The 
energy flux density needed to support the network chromosphere against 
radiation losses is  estimated to be 
a few times 
$10^7$ erg~cm$^{-2}$~s$^{-1}$ \citep{Anderson+Athay1989}.  The enthalpy flux density of
individual spicules with speeds of 100 \velu{} is $pv \sim 3\times
10^7$ erg~cm$^{-2}$~s$^{-1}$, which is thus comparable. 
Perhaps then these Hinode spicules are intrinsically related to the chromospheric
heating that is observed?

\section{Conclusions}

Hinode BFI Ca~II images obtained at the solar limb 
are consistent with the presence of the
stratified chromosphere when spicular emission is Doppler shifted
relative to the stratified material.  This can be achieved most naturally
using broad and/or Doppler shifted spicule line profiles of magnitudes
compatible with observed motions.  
The picture presented here can be tested directly using very stable
spectra at the solar limb, to see for example if the behavior modeled
in Figure~\pref{fig:standard} is qualitatively correct.  This is a
very challenging observation to make from the ground, but should be
possible under conditions of outstandingly good seeing and with modern
adaptive optics systems.  

The calculations reinforce a commonly known
problem regarding broad band spectral imagers: one must be very
careful taking care of physical effects such as Doppler motions which
are not spectrally resolved by the instrument.
BFI Ca II limb
observations are largely blind to the bulk of the chromosphere itself.
This fact is a sobering reminder that 
the absence of a signature can be as significant as its presence.

\figone
\figtwo
\figthree
\figfour
\figfive

\end{document}